\documentclass[structabstract]{aa}

\usepackage{txfonts}
\usepackage{graphicx}
\usepackage{natbib}
\bibpunct{(}{)}{;}{a}{}{,} 
\usepackage{color}

\newcommand{\ME}{M$_{\oplus}$ }

\newcommand{\MJ}{M$_{\mathrm{J}}$}

\begin{document}
 
\title{On the degeneracy of the tidal Love number $k_2$ in multi-layer planetary models: application to Saturn and GJ\,436b}

\author{U. Kramm\inst{\ref{UniHRO}}\thanks{\email{ulrike.kramm2@uni-rostock.de}} \and N. Nettelmann\inst{\ref{UniHRO},\ref{UCSC}} \and R. Redmer\inst{\ref{UniHRO}} \and D. J. Stevenson\inst{\ref{Caltech}}}

\institute{Institute of Physics, University of Rostock, D-18051 Rostock \label{UniHRO}
\and
Department of Astronomy and Astrophysics, University of California, Santa Cruz, CA 95064 \label{UCSC}
\and
Division of Geological and Planetary Sciences, California Institute of Technology, Pasadena, CA 91125 \label{Caltech}}

\date{Received 22 September 2010 / Accepted 23 December 2010}

\abstract
{In order to accurately model giant planets, a whole set of observational constraints is needed. As the conventional constraints for extrasolar planets like mass, radius, and temperature allow for a large number of acceptable models, a new planetary parameter is desirable in order to further constrain planetary models. Such a parameter may be the tidal Love number $k_2$.}
{In this paper we aim to study the capability of $k_2$ to reveal further information about the interior structure of a planet.}
{With theoretical planetary models we investigate how the tidal Love number $k_2$ responds to the internal density distribution of a planet. In particular, we demonstrate the effect of the degeneracy of $k_2$ due to a density discontinuity in the envelope of a three-layer planetary model.}
{The effect of a possible outer density discontinuity masks the effect of the core mass on the Love number $k_2$. Hence, there is no unique relationship between the Love number $k_2$ and the core mass of a planet. We show that the degeneracy of $k_2$ with respect to a layer boundary in the envelope also occurs in existing planets, e.g. \object{Saturn} and the Hot Neptune \object{GJ\,436b}. As a result of the degeneracy, the planetary parameter $k_2$ cannot be used to further constrain Saturnian models and for GJ\,436b only a maximum possible core mass can be derived from a given $k_2$. To significantly narrow the uncertainty about the core mass of GJ\,436b the combined knowledge of $k_2$ and atmospheric metallicity and temperature profile is necessary.}
{}


\titlerunning{Degeneracy of the tidal Love number $k_2$}
\authorrunning{Kramm et al.}

\maketitle

\section{Introduction}

Ever since the first extrasolar planet around a solar-type star was detected
\citep{MayorQueloz95}, questions about the composition and origin of extrasolar
planetary objects (exo-planets) have been of major interest. Models of
exoplanets are often little constrained based on the observable parameters mass,
radius, and effective temperature, in particular metal-rich planets
\citep{Adamsetal08}. For solar planets additional constraints are provided by
the gravitational moments which have been measured by spacecraft or Earth-based
observations of the motion of satellites and hence are not accessible for
extrasolar planets. However, a similar quantity does exist: the tidal Love
number $k_2$. To first order in the dimensionless number that describes the
effect of rigid rotation or degree 2 tidal distortion $k_2$ is equivalent to
$J_2$ \citep[see e.g.][]{Hubbard84}.

The tidal Love number $k_2$ is a potentially observable parameter.
\citet{RagozzineWolf09} showed that the dominant source of apsidal precession of
Hot Jupiters is the tidal interaction between the planet and its star.
This tidally induced apsidal precession creates a unique variation in the
transit light curve which is detectable by space-based missions like
\emph{Kepler}. Another possibility of determining $k_2$ is the measurement of
the orbital parameters of a two-planet system in apsidal alignment
\citep{Batyginetal09}. Due to tidal dissipation a coplanar two-planet system can
evolve into a tidal fixed point which is characterized by the alignment of the
apsidal lines \citep{Mardling07} and both orbits precess with the same rate.
\citet{Batyginetal09} showed that in this state the Love number $k_2$ is a
function of the inner planet's eccentricity.

Like $J_2$ for the solar system planets, $k_2$, if known, can be used to further
constrain the models of extrasolar planets as it is sensitive to the internal
density distribution of the planet. Understanding the planetary interior is
important for determining  not only physical processes but also the formation
history. Hence, it is crucial to analyze what information can be extracted from
a measured $k_2$ and its implications on the planetary interior.

First, we will describe the definition and calculation of the Love numbers in
Sect.~\ref{sec:DefCalc2L}. We also confirm the correlation between the central
condensation of a planet and its Love number $k_2$ within a simple
two-layer model. In Sect.~\ref{sec:3L-model} we introduce a more sophisticated
three-layer planetary model and demonstrate the degeneracy of $k_2$ with respect
to the density discontinuity in the envelope. We apply these results to Saturn
and to the Hot Neptune GJ\,436b in Sect.~\ref{sec:planets}. The main results of
this paper are summarized in Sect.~\ref{sec:sum}.

\section{The Love number $k_2$} \label{sec:DefCalc2L}

\subsection{Definition \& calculation}

Love Numbers quantify the deformation of the gravity field of a planet in response to an external perturbing body of mass $M$, which can be the parent star, another planet or a satellite. $M$, moving in a circular orbit of radius $a$ around a planet, causes a tide-raising potential \citep{ZharkovTrubitsyn78}
\begin{equation}
 W(s)=\sum^\infty_{n=2}W_n=(GM/a)\sum^\infty_{n=2}(s/a)^nP_n(\cos\theta')\quad, \label{eq:tidalpotential}
\end{equation}
where $s$ is the radial coordinate of the point under consideration inside the planet, $\theta'$ the angle between the planetary mass element at $s$  and the center of mass of $M$ at $a$, and $P_n$ are Legendre polynomials. Due to the tidally induced mass shift the planet's potential changes by $V^{\mathrm{ind}}_n(s)=K_n(s)W_n(s)$, where $K_n(s)$ is the Love function \citep{Love11}. Thus, at the planet's surface the definition of the Love numbers $k_n$ reads
\begin{equation}
 V^{\mathrm{ind}}_n(R_\mathrm{p})=k_nW_n(R_\mathrm{p})\quad.
\end{equation}

As we are interested in low eccentricity synchronous orbits, we
concentrate on the purely hydrostatic tides.
For the calculation of the Love numbers we follow the approach by \citet{ZharkovTrubitsyn78}, see also \citet{GavrilovZharkov77} and \citet{Gavrilovetal75}. A Love number of degree $n$ is obtained from
\begin{equation}
 k_n=\frac{T_n(R_\mathrm{p})}{R_\mathrm{p}g_0}-1 \quad,
\end{equation}
where $T_n(R_\mathrm{p})$ is the value of the function $T_n(s)$ at the planet's surface, $R_\mathrm{p}$ is the radius of the planet and $g_0$ the surface gravity for the unperturbed planet. The function $T_n(s)$ satisfies the following second order differential equation:
\begin{equation}
 T''_n(s)+\frac{2}{s}T'_n(s)+\left[\frac{4\pi G \rho'(s)}{V'(s)}-\frac{n(n+1)}{s^2}\right]T_n(s)=0 \quad.
\end{equation}
The radial coordinate is represented by $s$ and $\rho(s)$ and $V(s)$ give the unperturbed density distribution and potential of the planet, respectively. The primes denote first and second differentiation with respect to the radius $s$. If the planet has an internal density jump then the jump condition for the function $T$ is
\begin{eqnarray}
 T_n(b^+) = T_n(b^-)\quad,\\
 T'_n(b^+) = T'_n(b^-) + \frac{4\pi G}{V'(b)} \left[\rho(b^-)-\rho(b^+)\right]T_n(b) \quad. 
\end{eqnarray}
Here $b$ is the radial position of the density jump and the ''+'' (''-'') denotes a place just outside (inside) the discontinuity. This procedure assumes linear response, i.e. the tidal distortion is assumed to be small. Note that the \emph{only} input needed for the calculation of a Love number $k_n$ is the radial density distribution $\rho(s)$ of the planet. Hence, the Love numbers contain important information about the interior structure of a planet.
In this paper we focus on the Love number $k_2$ (twice the apsidal motion constant described by \citet{Sterne39}).

\subsection{Central condensation \& two-layer model}

The Love number $k_2$ is of special interest. It is a measure for the level of central condensation of an object: The more homogeneous the planet in mass distribution, the bigger the Love number $k_2$. Maximum homogeneity is represented by a planet of constant density, yielding the maximum value of $k_2$ of 1.5. If the planet is more centrally condensed, the Love number decreases. A planet with a density distribution of a $n=1$ polytrope has a value of $k_2=15/\pi^2-1=0.5198\dots$ Planets with a core can have an even smaller Love number due to a stronger central condensation, e.g. Saturn interior models with a $\sim10$\,\ME core give $k_2=0.32$. That is why it has been suggested to use $k_2$ to infer the presence of a massive core \citep{RagozzineWolf09}. 

First, we investigate the dependence of $k_2$ on the density distribution within a simple two-layer model, consisting of a core with constant density $A$ and an $n=1$ polytropic envelope:
\begin{eqnarray}
 \rho_1(x)=\frac{\sin q(1-x)}{qx} \quad,\quad x_\mathrm{c} \leq x \leq 1 \quad \label{eq:2L-modela}\\
 \rho_2(x)=A  \quad,\quad 0\leq x \leq x_\mathrm{c} \quad. \label{eq:2L-modelb}
\end{eqnarray}
The functions $\rho_1(x)$ and $\rho_2(x)$ give the radial density distribution of the envelope and core, respectively. The radius coordinate has been scaled such that $x=1$ is the surface of the planet and $x_\mathrm{c}$ gives the radius of the core in planet radii ($x=r/R_\mathrm{p}$). The core density $A$ is always greater than $\rho_1(x_\mathrm{c})$. The free parameters of this model are the core radius $x_\mathrm{c}$ and the quantity $q$. While $x_\mathrm{c}$ only influences the core density, $q$ influences both the core density and the density in the envelope. The core density $A$ is determined by the choice of $x_\mathrm{c}$ and $q$ under the condition that just outside of the core hydrostatic equilibrium is satisfied. Due to the strong influence of $q$ on the core density $A$, a change in $q$ is equivalent with changing the ratio of core mass to total mass $M_{\mathrm{core}}/M_{\mathrm{total}}$\footnote{$M_{\mathrm{core}}=4\pi A x_\mathrm{c}^3/3$ and $M_{\mathrm{total}}=M_{\mathrm{core}}+4\pi\int_{x_\mathrm{c}}^1 \rho_1(x)x^2\,\mathrm{d}x$}. It is necessary to choose $q\leq\pi$ since we must have $\rho_1(x)>0$. At the surface $\rho_1(1)=0$.

The use of an $n=1$ polytrope is motivated by two considerations. First, it is a good approximation for the behavior of hydrogen-helium mixtures for the range of pressures and temperatures of interest and is even still roughly correct if modest amounts of ice or rock are mixed with hydrogen and helium. Second, it yields an analytic form (given above) for the density profile, thus making the Love number calculations straightforward. The parameter $q$ is directly related to the proportionality constant $K$ in the assumed polytropic equation of state $P=K\rho^{(1+1/n)}$. Since our intent here is to understand the general nature of the dependence of the Love number on density structure and not to derive highly specific and precise models, it serves our purpose well.

\begin{figure}
 \resizebox{\hsize}{!}{\includegraphics{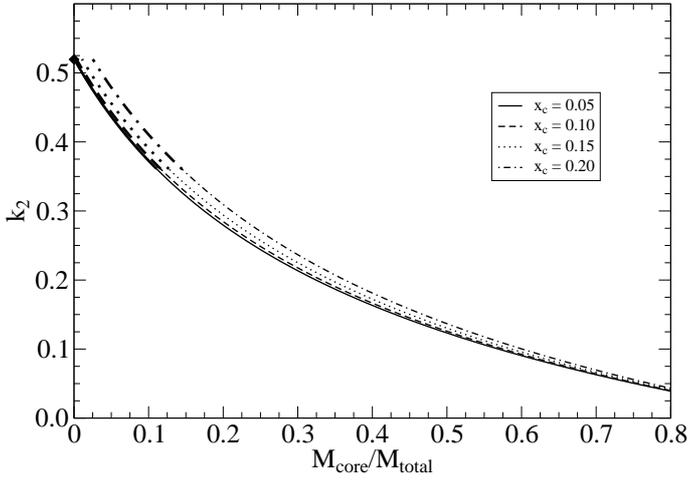}}
 \caption{Love number $k_2$ in dependence on the ratio of core mass to total mass for theoretical two-layer models consisting of a constant density core and an $n=1$ polytropic envelope, see equations (\ref{eq:2L-modela}) and (\ref{eq:2L-modelb}). The lines show solutions for core radii of $x_\mathrm{c} = 0.05$ (solid), 0.10 (dashed), 0.15 (dotted) and 0.20 (dash-dotted). The black diamond shows the solution for the core-less polytropic planet. To obtain different values for $M_{\mathrm{core}}/M_{\mathrm{total}}$ the parameter $q$ has been varied. $q=\pi$ gives the minimum value of $M_{\mathrm{core}}/M_{\mathrm{total}}$. Thick lines show the region where $0.9\pi < q < \pi$ which corresponds to realistic $\rho_1(x_\mathrm{c})/A$ ratios. }
 \label{fig:2L_massratio_k2}
\end{figure}

By choosing $x_\mathrm{c}$ and $q$ we can construct different two-layer models. Figure~\ref{fig:2L_massratio_k2} shows the Love number $k_2$ as function of $M_{\mathrm{core}}/M_{\mathrm{total}}$ for different core radii of respectively $x_\mathrm{c}=0.05$, 0.10, 0.15 and 0.20, and in the core-less case. To obtain different values for $M_{\mathrm{core}}/M_{\mathrm{total}}$ the parameter $q$ has been varied. 
A value of $q=\pi$ gives the minimum ratio $M_{\mathrm{core}}/M_{\mathrm{total}}$ as the core density $A$ decreases with $q$. Reducing the $q$ parameter causes a rapid increase of the core density $A$ but only a modest increase of $\rho_1(x_\mathrm{c})$. For the $q$ parameter range of $0.9\pi < q < \pi$ the ratio of the envelope density at the core-mantle boundary and the core density has values of $\rho_1(x_\mathrm{c})/A \gtrsim 0.01$. Models with too low $q$ become unrealistic as they have an enormously small $\rho_1(x_\mathrm{c})/A$. Note that the choice of the core radius $x_\mathrm{c}$ has a compensating effect and $q$ could be choosen lower if $x_\mathrm{c}$ is bigger. In reality, lower $k_2$ can be obtained in particular if the mantle density distribution does not follow an $n=1$ polytrope and a compressive equation of state for the core material is used.
As can be seen from Fig.~\ref{fig:2L_massratio_k2}, $k_2$ decreases with increasing ratio of core mass to total mass. This behavior applies for all choosen core radii. For a planet of a given ratio of core mass to total mass, a bigger core radius $x_\mathrm{c}$ implies a lower density for that core and hence less central condensation and a larger Love number, as Fig.~\ref{fig:2L_massratio_k2} shows. The data points for the smallest mass ratios have been generated with the maximum value of $q=\pi$. This means that for a two-layer model with a fixed core radius there is a minimum core mass.

These calculations within a simple two-layer model give an intuitive interpretation of the characteristic of $k_2$ of being a measure for the level of central condensation: the bigger the core mass, the more centrally condensed and hence the smaller the Love number $k_2$. However, as we will show in the next section this simple deduction is no longer possible when there is another density discontinuity in the envelope of the planet.

\section{Degeneracy in three-layer models} \label{sec:3L-model}

In this section we will investigate the behavior of $k_2$ in a more complicated model. In addition to the density discontinuity at the core-mantle boundary we introduce another discontinuity in the envelope of the planet. A three layer structure is a common assumption in planet modeling and has been used for modeling the solar system giants Jupiter and Saturn (see e.g. \citet{Guillot99}, \citet{SaumonGuillot04}). Such a separation of layers in the planetary envelope can occur as a result of demixing of hydrogen and helium \citep{StevensonSalpeter77}. It could also arise from double diffusive convection in the presence of a density gradient that is introduced during accretion or because of subsequent core erosion \citep{Stevenson82}. Layer boundaries are also compatible with standard models of planet formation \citep{Hubbardetal95}. We define our theoretical three-layer model as follows:
\begin{eqnarray}
 \rho_1(x)=\frac{\sin q_1(1-x)}{q_1x} \quad,\quad x_\mathrm{m} \leq x \leq 1 \label{eq:3L-modela}\\
 \rho_2(x)=B\frac{\sin q_2(x_\mathrm{a}-x)}{q_2 x}  \quad,\quad x_\mathrm{c} \leq x \leq x_\mathrm{m} \label{eq:3L-modelb}\\
 \rho_3(x)=A \quad,\quad 0 \leq x \leq x_\mathrm{c} \quad. \label{eq:3L-modelc}
\end{eqnarray}

It consists of a core with constant density $A$ and two polytropic envelopes described by $\rho_1(x)$ (outer envelope) and $\rho_2(x)$ (inner envelope). The same characteristics as for the two-layer model apply. The layer boundary in the envelope is placed at $x_\mathrm{m}$. We choose the parameters $q_1$ and $q_2$ (near but smaller than $\pi$ and not much different from each other). The location of the density discontinuities $x_\mathrm{c}$ and $x_\mathrm{m}$ are also free parameters. The core density $A$ is determined as for the two-layer model. The continuity of pressure and gravity at $x_\mathrm{m}$ is used to calculate $x_\mathrm{a}$. We can then choose $B$ to get a specified non-dimensionalized size of the density jump $\Delta\rho=\left.(\rho_2-\rho_1)/\rho_1\right|_{x=x_\mathrm{m}}$. Summarizing, in this three-layer model we can vary the parameters $x_\mathrm{c}$, $x_\mathrm{m}$, $q_1$, $q_2$ and $\Delta\rho$ while $A$, $B$ and $x_\mathrm{a}$ are determined by the parameters chosen.

\subsection{Love number and core mass} \label{subsec:LNandCM}

The parameters characterizing the density discontinuity in the envelope are $x_\mathrm{m}$ and $\Delta\rho$, position and size of the discontinuity. Thus, in order to investigate the influence of the outer discontinuity on the Love number $k_2$ we varied the parameters $x_\mathrm{m}$ and $\Delta\rho$ from 0.5 to 0.9 planet radii (increment 0.01) and from 0.01 to 0.5 (increment 0.01), respectively. For the example we give here, the other free parameters are fixed at $q_1=0.98\pi$, $q_2=0.99\pi$ and $x_\mathrm{c}=0.1$. This choice is in order to keep the envelope structure close to an $n=1$ polytrope. Together with a moderate value for the core size this example for the density distribution mimics a Jupiter-like planet.

\begin{figure}
 \resizebox{\hsize}{!}{\includegraphics{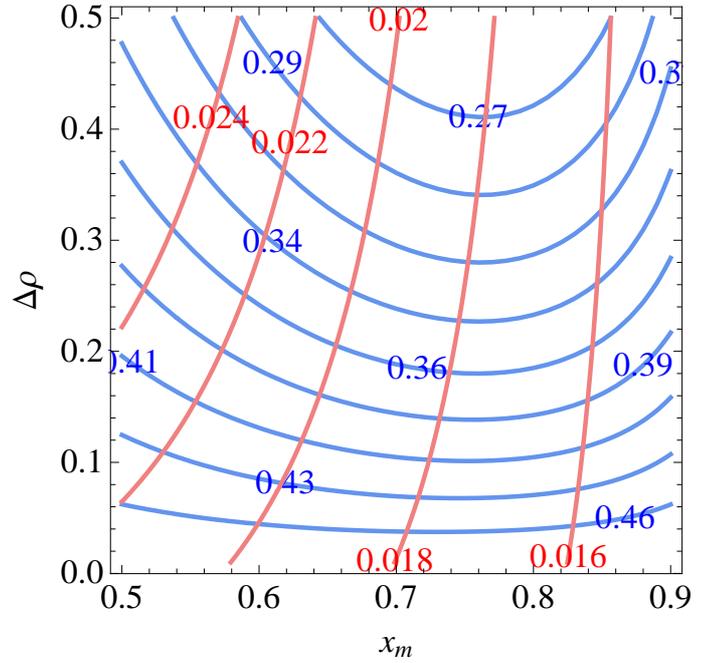}}
 \caption{Lines of equal $k_2$ (blue) and equal $M_{\mathrm{core}}/M_{\mathrm{total}}$ (red) in $\Delta\rho$-$x_\mathrm{m}$ parameter space for a theoretical three-layer model, see equations (\ref{eq:3L-modela}), (\ref{eq:3L-modelb}) and (\ref{eq:3L-modelc}) with the fixed parameters $q_1=0.98\pi$, $q_2=0.99\pi$ and $x_\mathrm{c}=0.1$. Numbers give the corresponding values of $k_2$ and $M_{\mathrm{core}}/M_{\mathrm{total}}$, respectively. The equi-$k_2$-lines demonstrate the degeneracy of $k_2$ with respect to the position ($x_\mathrm{m}$) and size ($\Delta\rho$) of a density discontinuity in the envelope and the ratio of core mass to total mass (a blue line intersects several red lines).}
 \label{fig:3L-degeneracy}
\end{figure}

The change of the Love number $k_2$ in dependence on the parameters of the outer density discontinuity is shown by Fig.~\ref{fig:3L-degeneracy}. Lines of equal $k_2$ and their values are given. This demonstrates a \emph{degeneracy of the Love number $k_2$ with respect to the outer density discontinuity}. One can always find many different $x_\mathrm{m}$-$\Delta\rho$-pairs (that means different three-layer planetary models) that give the same $k_2$. These models lie on one of the equi-$k_2$-lines and hence one cannot distinguish between these models by a measurement of $k_2$. 
In addition to the Love number $k_2$ we calculated the ratio of core mass to total mass $M_{\mathrm{core}}/M_{\mathrm{total}}$ in the same parameter space. Lines of equal mass ratio are also shown in Fig.~\ref{fig:3L-degeneracy}. For a specific value of $k_2$ planetary models with different mass ratios are possible, see intersections of equi-$k_2$- and equi-$M_{\mathrm{core}}/M_{\mathrm{total}}$-lines. (Note that the resulting $M_{\mathrm{core}}/M_{\mathrm{total}}$ have a very limited range because only $x_\mathrm{m}$ and $\Delta\rho$ change and $M_{\mathrm{core}}/M_{\mathrm{total}}$ is more sensitive to the $q$ parameters which are fixed in our example here. Anyhow, other choices of the $q$ parameters and/or $x_\mathrm{c}$ would yield qualitatively similar results.) In the same manner, planetary models with identical mass ratios can have different $k_2$ values. This demonstrates that a unique relationship between the core mass and the Love number $k_2$ is no longer valid in a three-layer model. It can also be seen from Fig.~\ref{fig:3L-degeneracy} that planets with a more massive core can have a bigger Love number than planets with a smaller core (compare e.g. left lower part with right upper part of Fig.~\ref{fig:3L-degeneracy}). This is in contrast to the results that were suggested by the simple two-layer model where more massive cores lead to smaller $k_2$. In conclusion, the effect of a possible density discontinuity in the envelope of a planet on the Love number $k_2$ can mask the effect of the core mass.

Even though we found that there is a non-unique relation between the Love number $k_2$ and the core mass, Fig.~\ref{fig:3L-degeneracy} is still consistent with the assumption of $k_2$ being a measure for the level of central condensation of a planet as the term 'central condensation' cannot be defined by the core mass alone. A big density jump $\Delta\rho$ means that there is a significantly higher density in the inner than in the outer envelope. This constitutes a higher level of central condensation than an envelope without a discontinuity. That is why $k_2$ is generally smaller in regions with bigger $\Delta\rho$. On the other hand, a very small $\Delta\rho$ means that there is a small difference in density at the layer boundary\footnote{The limiting case $\Delta\rho=0$ is equal with a two-layer model like it was discussed above.}. Hence, the planet is more homogeneous and the Love number $k_2$ tends to bigger values. In the region of small $\Delta\rho$ the equi-$k_2$ lines also become flatter, because the position $x_\mathrm{m}$ of the discontinuity becomes less important for $\Delta\rho\rightarrow0$, as the distinguishability between a two-layer and a three-layer model disappears.

\subsection{Moment of inertia}

Another quantity closely related to the structure of a planet is its
moment of inertia $C$, or its non-dimensional form
$C_\mathrm{nd}=C/MR_\mathrm{p}^2$. While it has been possible to obtain moments
of inertia from measurements of precession for Earth and Mars \citep[see
e.g.][]{Folkneretal97}, it is more difficult for the giant planets. For the
modeling of the internal structure of the giants, the gravitational moments have
been used instead. The first-order response of a planet to rotational distortion
(quantified by $J_2$) can be related to the planet's nondimensional polar moment
of inertia with the Radau-Darwin equation \citep{Hubbard84}:
\begin{equation}
 C_{\mathrm{nd}}^\mathrm{(RD)}=\frac{2}{3}\left[1-\frac{2}{5}\left(\frac{5}{3\Lambda_{2,0}^\mathrm{(r)}+1}-1\right)^\frac{1}{2}\right] \quad,
 \label{eq:RD}
\end{equation}
where $M$ and $R_\mathrm{p}$ are the mass and the radius of the planet,
respectively, and to linear order $\Lambda_{2,0}^\mathrm{(r)}=J_2/q$ with
$q=\omega^2R_\mathrm{p}^3/GM$, the ratio of centrifugal force to gravity. This
equation implies that the response coefficient $\Lambda_{2,0}^\mathrm{(r)}$ is a
function of the moment of inertia only.

Since the tidal potential~(\ref{eq:tidalpotential}) has the same
functional form as the rotational potential, the tidal response of a planet can
be treated analogously to the rotational one. For a liquid or perfectly elastic
planet the tidal response coefficient $\Lambda_{2,0}^\mathrm{(t)}=k_2/3$ is the
same as the rotational one \citep{Hubbard84}. Substituting
$\Lambda_{2,0}^\mathrm{(t)}$ for $\Lambda_{2,0}^\mathrm{(r)}$ in
equation~(\ref{eq:RD}), we find:
\begin{equation}
 C_{\mathrm{nd}}^\mathrm{(RD)}=\frac{2}{3}\left[1-\frac{2}{5}\left(\frac{5}{k_2+1}-1\right)^\frac{1}{2}\right]\quad.
 \label{eq:RDk2}
\end{equation}

With this relation we can calculate the nondimensional moment of inertia
from our previously generated values of the Love number $k_2$. Another, more
direct, method to obtain the moment of inertia for our models is to integrate
over the density distribution:
\begin{equation}
 C_{\mathrm{nd}}=\frac{\frac{2}{3}\int_0^1x^4\rho(x)\,\mathrm{d}x}{\int_0^1x^2\rho(x)\,\mathrm{d}x}\quad.
 \label{eq:Crho}
\end{equation}

\begin{figure}
 \resizebox{\hsize}{!}{\includegraphics{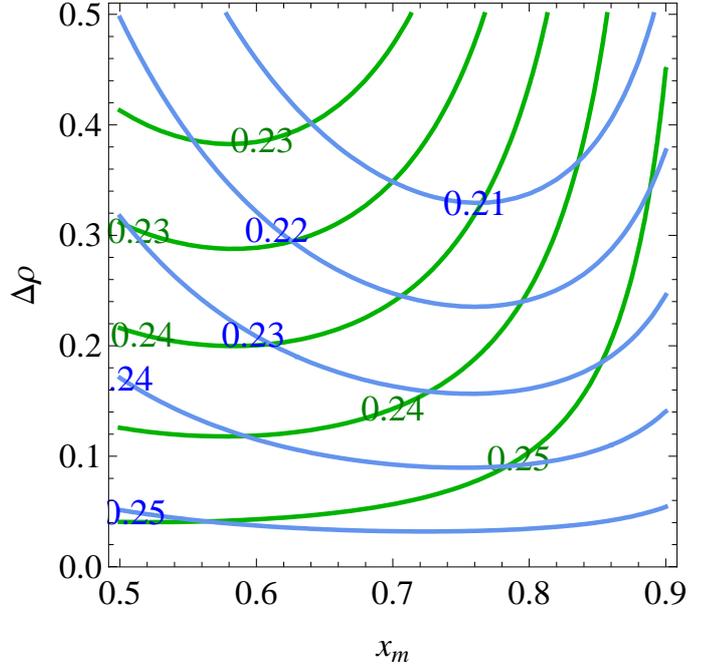}}
 \caption{Lines of equal moment of inertia $C_{\mathrm{nd}}$ for the same three-layer model as in~\S\ref{subsec:LNandCM} and in the same $\Delta\rho$-$x_\mathrm{m}$ parameter space. Blue lines are for the moments of inertia obtained directly from the Love numbers $k_2$ via equation~(\ref{eq:RDk2}). Green lines show the moments of inertia calculated from the density distribution with equation~(\ref{eq:Crho}).}
 \label{fig:MomentsOfInertia}
\end{figure}

We calculated moments of inertia for the reference three-layer model of
\S\ref{subsec:LNandCM}. In analogy to Fig.~\ref{fig:3L-degeneracy} the
characteristics of the envelope layer boundary are varied.
Fig.~\ref{fig:MomentsOfInertia} shows lines of equal moment of inertia, obtained
from equation~(\ref{eq:RDk2}) or~(\ref{eq:Crho}).

Of course the equi-$C_{\mathrm{nd}}$-lines generated from (\ref{eq:RDk2})
are equivalent to the equi-$k_2$-lines of Fig.~\ref{fig:3L-degeneracy}, as
(\ref{eq:RDk2}) is a one-to-one correspondence. Obviously, they are different
from the results obtained from~(\ref{eq:Crho}). This is due to the
approximations made for the derivation of equation~(\ref{eq:RDk2}). 

The Radau-Darwin equation~(\ref{eq:RD}) itself is an approximation. It
only approaches a correct result $C_\mathrm{nd}\rightarrow0.4$ for the limit of
a completeley homogeneous planet. It is not true in the limit
$C_\mathrm{nd}\rightarrow0$ \citep[see e.g.][]{MurrayDermott99}. This feature of
the Radau-Darwin equation can be seen in Fig.~\ref{fig:MomentsOfInertia}. The
difference between the equi-$C_\mathrm{nd}$-lines becomes bigger with increasing
$\Delta\rho$, corresponding to stronger central condensation.

We conclude that the Radau-Darwin relation is a too crude approximation
to describe the moment of inertia of gas giant planets.

\section{Application to existing planets} \label{sec:planets}

In this section we apply the results obtained for the simple models shown above to interprete interior models for real planets. We will show that the degeneracy of $k_2$ is not just an artifact of our theoretical three-layer model but actually occurs in existing solar as well as extrasolar planets.

\subsection{Saturn}

As the second largest planet in our solar system, Saturn has attracted much attention in the past. With the help of space missions like e.g. Cassini observational data have been derived that provides important constraints for modelling Saturn.
Voyager measurements have yielded a depletion of helium compared to the solar abundance \citep{ConrathGautier00}. In conclusion, the missing helium must be hidden deep inside the planet. The helium depletion in the outer part of the envelope can be explained by a phase separation of hydrogen and helium. \citet{Lorenzenetal09} and \citet{Moralesetal09} showed recently that demixing occurs on large scale at standard Saturn interior model conditions, motivating a layered-envelope assumption for Saturnian models. Furthermore, the demixing theory allows to reproduce the correct age of Saturn which was not possible without demixing \citep{FortneyHubbard03}.

As Saturn can be modeled in a three- or more layer approach, it is also affected by the $k_2$-degeneracy. We demonstrate this effect by examining selected Saturn models from \citet{Nettelmann09}. Those models have a rocky core based on the rock equation of state (EOS) by \citet{HubbardMarley89} and two isentropic envelopes of hydrogen, helium and water (as representative for metals) based on the linear mixing Rostock equation of state (LM-REOS), see \citet{Holstetal08}, \citet{Kietzmannetal07} and \citet{Frenchetal09}, respectively. The pressure at the layer boundary in the envelope is called the transition pressure $P_{12}$. By changing $P_{12}$ the layer boundary can be shifted.

The variation of $P_{12}$ affects the core mass of Saturn. We assumed a layer boundary between 1 and 3.8\,Mbar, where $P_{12}=1\,\mathrm{Mbar}$ gives a rocky core mass of 14\,\ME and $P_{12}=3.8\,\mathrm{Mbar}$ gives the maximum value of $P_{12}$ as the core vanishes when the layer boundary is put that deep inside the planet \citep{Nettelmann09}. It is important to note that all these models reproduce the observational constraints, including the gravitational moments $J_2$, $J_4$ and $J_6$. This means that the parameter $P_{12}$ is not fixed. As a consequence, there is a large uncertainty in the core mass of at least 14\,\ME. We calculated the Love number $k_2$ for these models and found for all these Saturn models $k_2\simeq0.32$. Even though the core mass varies fom 0 to 14\,\ME, no significant difference in the $k_2$ values can be seen. This is an expected result because $k_2$ is proportional to $J_2$ to first order in the expansion of the planet's potential \citep[see e.g.][]{Hubbard84} and all models match the observed $J_2$.

The constant $k_2$-value demonstrates the effect described in the previous section: The influence of the outer density discontinuity masks the effect of the core mass. For one $k_2$ value different planetary models can be found. Increasing the transition pressure $P_{12}$ is equivalent to shifting the density discontinuity inwards to smaller $x_\mathrm{m}$ (compare theoretical three-layer model in section~\ref{sec:3L-model}). At the same time the size of the density jump $\Delta\rho$ grows in order to obtain models that are consistent with  the measured gravitational moments. So all Saturn models presented in this work lie on an equi-$k_2$-line (compare Fig.~\ref{fig:3L-degeneracy}), showing the degeneracy of the Love number $k_2$ with respect to the outer layer boundary in Saturn. 

Figure~\ref{fig:3L-degeneracy} also predicts a growing core mass with a deeper layer boundary and increasing density jump. However, this is not applicable here because our Saturn models are based on the additional constraints of matching the gravitational moments. These constraints were not considered for the simple models in Sect.~\ref{sec:3L-model}. In the planet modelling procedure the gravitational moments are adjusted by varying the envelope metallicities. This means that the metallicities in the envelope are determined by the observed gravitational moments for a specified $P_{12}$. Shifting the layer boundary inwards (bigger $P_{12}$, smaller $x_\mathrm{m}$) forces the metallicity of the inner envelope to increase in order to ensure consistency with the gravitational moments. These ''additional'' metals for the inner envelope are taken from the core, which is why the core mass shrinks with increasing transition pressure~$P_{12}$.

\subsection{GJ\,436b}

For modelling extrasolar planets the Love number $k_2$ can be a valuable constraint because the gravitational moment $J_2$ is not accessible to observations. To first order $k_2$ is equivalent to $J_2$ \citep[see e.g.][]{Hubbard84} and is \emph{potentially observable} due to apsidal precession of the planet's orbit \citep{RagozzineWolf09} or measurement of the orbital parameters of a two-planet system in apsidal alignment \citep{Mardling07}. However, interior models of extrasolar planets are also subject to $k_2$-degeneracy. Here we demonstrate this effect with the example of the Hot Neptune GJ\,436b. Numerous investigations about this transiting planet have been made in the past \citep[see e.g.][]{Adamsetal08,Figueiraetal09,Torresetal08}. Based on mass, radius, and temperature measurements alone, a large variety of models is possible, ranging from a ''water world'' to a Super-Earth.
We have performed extensive calculations for modelling the interior structure of GJ\,436b \citep{Nettelmannetal10}. We considered two-layer models with a rocky core and one envelope where water is homogeneously mixed into H/He as well as three-layer models consisting of a rocky core, a water layer and an outer H/He layer with metal (water) abundance $Z_1$. For selected models we calculated the Love number $k_2$, for details about the modelling see \citet{Nettelmannetal10}. Here we present a new and extended version of Fig. 2 in \citet{Nettelmannetal10} in order to show the degeneracy of $k_2$.

\begin{figure}
 \resizebox{\hsize}{!}{\includegraphics{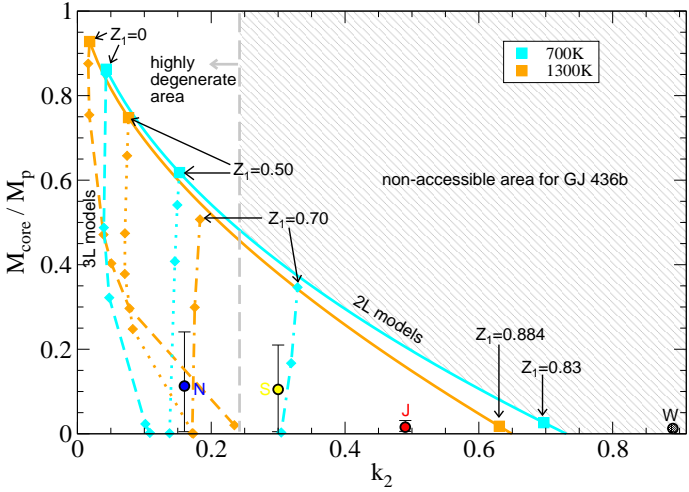}}
 \caption{Love number $k_2$ and core masses of two-layer models (squares) and three-layer models (diamonds) of GJ\,436b for temperatures of 700\,K (cyan) and 1300\,K (orange) of the isothermal (1-100\,bar) atmosphere. Color coded circles denote models of Jupiter (red), Saturn (yellow), Neptune (blue) and a 20\,\ME water planet (black). Three-layer models with the same metal abundance in the outer envelope are connected with dashed ($Z_1=0$), dotted ($Z_1=0.5$) and dash-dotted ($Z_1=0.7$) lines. The grey dashed line marks a highly degenerate area constructed to contain all $Z_1=0$ three-layer models (the position of this line has been shifted compared to Fig. 2 in \citet{Nettelmannetal10} in order to account for the new 1300\,K model series).}
 \label{fig:GJ436b}
\end{figure}

Figure~\ref{fig:GJ436b} shows Love numbers $k_2$ and core masses of various two- and three-layer models of GJ\,436b. All models shown here have an isothermal outer layer from 1 to 100\,bar. For comparison, Fig.~\ref{fig:GJ436b} also contains the solar planets Jupiter, Saturn and Neptune and a 20\,\ME water planet. For two-layer models solutions are shown which have a metal abundance of $Z_1=0$, 0.5, and the largest possible value when $M_{\mathrm{core}}=0$. Model series with three layers are displayed for constant outer envelope metallicities of $Z_1=0$, 0.5, and 0.7. Solutions are shown for two different surface temperatures of 700 and 1300\,K. Even though possible, we consider colder models to be unrealistic as recent results by \citet{Spiegeletal10} and \citet{Lewisetal10} predict 1300\,K and 1100\,K, respectively. In such warm models water is in the plasma phase and miscible with H/He. Hence, the water EOS applied to the inner envelope of such warm three-layer models represents a mean density of a H/He/H$_2$O/rock mixture.

For GJ\,436b we find $k_2=0-0.72$.
Models of GJ\,436b with $k_2=0.02-0.24$ can have any core mass between 0 and $\sim0.9$ planet masses. This means that a measurement of $k_2$ in this regime would not help constraining the interior further. This highly degenerate area includes all $Z_1=0$ three-layer models (note that this area has been extended compared to Fig. 2 in \citet{Nettelmannetal10} in order to account for the new 1300\,K model series).
However, one can predict maximum possible core masses for specified $k_2$. The uncertainty in the core mass decreases with increasing $k_2$. The upper limits of possible core masses, if $k_2$ and the atmospheric temperature profile are known, are necessarily given by two-layer models with one homogeneous envelope. Any redistribution of metals from an outer to an inner part of the envelope would, if the central condensation is to be kept at constant $k_2$ value, require a compensating transport of core material to the envelope. The core mass would therefore decrease and drop below these upper limits.
This is shown by the three-layer models with constant envelope metallicity of respectively $Z_1=0$, 0.5, 0.7, which all have lower core masses than the two-layer models. Three-layer models contain a density discontinuity in the envelope, dividing it into an inner water envelope ($Z_2=1$) and an outer H/He/H$_2$O envelope ($Z_1=0$, 0.5 and 0.7). The three-layer model approach gives the additional free parameter $P_{12}$, which is the pressure at the envelope layer boundary. To construct the equal-metallicity lines the metallicities are kept at constant values ($Z_2=1$ and $Z_1=0$, 0.5, 0.7) and $P_{12}$ is varied from a minimum value where $M_{\mathrm{core}}=0$ to a maximum value where the water layer vanishes and the model is identical to a two-layer model with the same envelope metallicity (for the modelling procedure see also \citet{Nettelmannetal10}). 

Models of equal envelope metallicity $Z_1$ show systematic behavior in $M_{\mathrm{core}}$-$k_2$ space. For low outer envelope metallicities the Love Number $k_2$ shrinks with growing core mass due to the stronger central condensation created by the bigger core. With a high metal abundance in the outer envelope this behavior changes and $k_2$ increases slightly with growing core mass. The higher amount of metals in the outer envelope leads to a smaller density gap at the envelope layer boundary so that the planet becomes more homogeneous than planets with a low metallicity envelope. As can be seen from Fig.~\ref{fig:GJ436b}, this effect, and especially the transition between the two different slopes, is strongly temperature dependent. For lower temperatures (see 700\,K curves) $k_2$ increases with the core mass already at smaller metallicities than for warmer temperatures (compare 1300\,K curves). That is because colder envelope temperatures increase their density and hence the homogeneity of the planet.
The upper limit of possible core masses increases with decreasing atmospheric temperatures. 

Concluding, an observational $k_2$ would imply a maximum possible core mass. But, especially for $k_2<0.24$, it would not help to further constrain interior models of GJ\,436b because in that regime the solutions are too degenerate. However, a $k_2>0.24$, if measured, would indicate a maximum core mass $M_\mathrm{c}<0.5\,\mathrm{M}_\mathrm{p}$ and large outer envelope metallicities.
Only with additional information about the planet's atmospheric composition ($Z_1$) together with $k_2$ interior models of GJ\,436b can be further constrained. If we had knowledge about the atmospheric temperature profile \emph{and} metallicity \emph{and} the Love number $k_2$, the core mass could be determined or at least strongly narrowed, depending on the slope of the equi-metallicity-lines. However, not all aspects of metallicity are potentially measurable in an exoplanet, only those that are the consequence of elements which are in the gas phase in the observable part of the atmosphere. As an example in our own solar system, the metallicities of Jupiter and Saturn are not measurable spectroscopically because water forms clouds deep below the photosphere.

The results for GJ\,436b are in perfect agreement with our results obtained from Fig.~\ref{fig:3L-degeneracy}. For a given $k_2$ value several planetary models and core masses are possible (highly degenerate area in Fig.~\ref{fig:GJ436b}). Even for constant outer envelope metallicity there are also models that have a larger $k_2$, indicating a rather homogeneous planet, but nevertheless a more massive core. This behavior is caused by the outer density discontinuity at the water-H/He layer boundary as discussed in Sect.~\ref{sec:3L-model} and predicted by Fig.~\ref{fig:3L-degeneracy}.

\section{Summary} \label{sec:sum}

In this paper we investigated the effect of the density distribution of a planet
on its tidal Love number $k_2$ in order to find out what conclusions can
be drawn from a measured $k_2$ on the internal structure of a planet. We
confirmed that the Love number $k_2$ is a measure of the level of central
condensation of a planet. However, in a three layer model approach $k_2$ is
\emph{not} a unique function of the core mass. There is a degeneracy of $k_2$
with respect to a density discontinuity in the envelope of the planet. It is
possible to have several acceptable models for a given $k_2$ value, which can
differ significantly in core mass. The effect of the outer density discontinuity
on $k_2$ is compensating the effect of the core. Furthermore, we showed
that the Radau-Darwin relation is a too crude approxomation to describe the
moment of inertia of gas giant planets.

We verified our results on $k_2$ with models of existing planets. For
Saturn the freedom to place the layer boundary in the envelope leads to a high
uncertainty in the core mass. Regardless of the core mass all Saturn models have
the same $k_2$, demonstrating the degeneracy caused by the outer layer boundary.

For extrasolar planets the Love number $k_2$ can be an equivalent constraint to $J_2$ for the solar system planets. However, one has to be careful with estimates about the core mass derived from $k_2$ as degeneracy may also occur in extrasolar planets. For GJ\,436b we find a highly degenerate area of $k_2<0.24$ where a measurement of $k_2$ would barely help to further constrain the interior models. Only a \emph{maximum possible} core mass and for $k_2>0.24$ a large metallicity can be inferred. With additional knowledge about the atmospheric metal abundance the uncertainty about the core mass could be significantly narrowed.

Tabulating $k_2$ values of various planetary models can prove to be very useful once $k_2$ is actually measured for extrasolar  transiting planets. For instance, for the Super-Earth \object{GJ\,1214b} \citet{Nettelmannetal10b} demonstrated that H/He or water envelopes result in significantly different values of $k_2$.
Furthermore, we have shown in this paper that even though the Love number $k_2$ is a degenerate quantity it can help constraining the core mass of a planet. Knowledge about the core masses of planets is highly desired because it is thought to help to distinguish between the possible planet formation scenarios of core accretion \citep[see e.g.][]{Pollacketal96} and gravitational instability \citep[see e.g.][]{Boss97}. However, one has to keep in mind that core accretion models can also result in very small cores of 1.7\,\ME or 0.25\,\ME in the case of grain-free or even metal-free envelopes, respectively \citep{HoriIkoma10}. On the other hand, gravitational instability models allow the formation of a massive core as well if the protoplanet is cold enough for grain settling to take place \citep{HelledSchubert08}. A clear distinction between the two formation models can only be made for massive extra-solar giant planets $\geq$ 5\,\MJ. \citet{HelledSchubert08} showed that such massive protoplanets formed by disk instability cannot build up a core at all due to their high internal temperatures and evaporation of the grains.

\begin{acknowledgements}
We acknowledge helpful discussions with Ralph Neuh\"auser. UK and RR acknowledge support from the DFG SPP 1385 ''Young Planets''.
\end{acknowledgements}

\appendix

\bibliographystyle{aa} 
\bibliography{15803} 

\end{document}